\documentclass[ nofootinbib, preprint]{revtex4-1}
\usepackage{amsmath,amssymb,pxfonts}
\usepackage{verbatim,graphicx}

\usepackage{color}
\usepackage[all]{xy}

\newcommand{\vev}[1]{\langle #1\rangle}
\newcommand{\ket}[1]{| #1\rangle}

\newcommand{\braket}[2]{\langle #1|#2\rangle}
\newcommand{\BR}{\mathbb{R}}

\newcommand{\CZ}{{{\cal Z}}}

\newcommand{\diag}{\hbox{diag}}

\def\Tr{\textrm{Tr}}

\newcommand{\beq}{\begin{equation}}
\newcommand{\beqs}{\begin{equation*}}
\newcommand{\eeq}{\end{equation}}
\newcommand{\eeqs}{\end{equation*}}

\allowdisplaybreaks

\begin{document}
\setlength{\unitlength}{1mm}
\title{On the central charge extension of the ${\cal N}=4 $  SYM spin chain}

\author{David Berenstein}
\affiliation { Department of Physics, University of California at Santa Barbara, CA 93106\\\\
and Department of Applied Math and Theoretical Physics, Wilbeforce Road, Cambridge, CB3 0WA, UK}

\begin{abstract} 
In this paper it is argued that the central charge extension of the Coulomb branch of ${\cal N}=4 $ SYM theory appears as a limit of Beisert's central charge extension of the planar ${\cal N}=4 $ spin chain in the presence of boundaries. These boundaries are interpreted as D-branes that source the central charge and are  realized as giant gravitons and dual giant gravitons in the AdS dual. The BPS states that correspond to short representations of the centrally extended algebra on the spin chain can stop from existing when they cross walls of stability that depend on the position of the branes. These walls can be understood easily at weak coupling in the $SU(2)$ sector.
\end{abstract}

\preprint{DAMTP-2014-85}

\maketitle

\section{Introduction}
\label{S:Introduction}

An amazing fact of ${\cal N}=4 $ SYM is that the planar diagrams lead to an integrable model that computes anomalous dimensions of single trace operators (see \cite{Beisert:2010jr} for a detailed review). It is standard to think about this integrable model as a set of defects scattering from each other, defects that are interpreted as  excitations of an infinitely long ferromagnetic ground state which preserves half the supersymmetries (this is an asymptotic Bethe ansatz \cite{Beisert:2005fw}). The theory then receives finite size corrections when the chain becomes finite in length rather than infinitely long.
This is a rather natural geometric starting point of view considering the Penrose limit in the $AdS_5\times S^5$ dual geometry \cite{Berenstein:2002jq}. 
 
Within this paradigm one can consider an infinite spin chain limit and study the properties of these excitations individually. Beisert \cite{Beisert:2005tm} has argued that in this limit the $PSU(2,2|4)$ symmetry algebra of the original theory is broken to a $SU(2|2)\times SU(2|2) \lJoin \BR$ subgroup, and that moreover the excitations carry the quantum numbers of a central extension of this subalgebra. 
The central charge essentially measures the quasi-momentum of the excitation on the chain. The central charge describes excitations even though the original $PSU(2,2|4)$ does not admit a central extension. The way this works out is that for a closed string the net central charge vanishes (is confined) due to the level matching constraints. From this point of view the additional central charge seems like an advantageous auxiliary construction. 

Consider now a theory of open superstrings in flat Minkowski space. This requires the introduction of D-branes \cite{Dai:1989ua,Polchinski:1995mt}. The lowest lying string modes stretching between two such D-branes (if they are flat, parallel and separated from each other) correspond to a massive short representation of the unbroken supersymmetry of the D-brane configuration. These representations require a central charge extension of the unbroken supersymmetry algebra in order to be short, but none of the closed strings will carry that central charge. The central charge only becomes physical in the open string sector (or when we compactify the theory on a circle), and this can be argued to be related to additional topological charge \cite{Witten:1978mh}. The additional charge is the electric charge carried by the end-points of the string, which can only become measurable in the field theory limit when we spontaneously break the non-abelian gauge symmetry on the stack of the branes. This corresponds to the Coulomb branch of the maximally supersymmetric  Yang Mills  theory on the world volume of the D-branes. From here on we will assume that these D-branes have a $3+1$ dimensional worldvolume, as these are the only examples that arise in the context of this paper. In this sense they have ${\cal N}=4 $ Supersymmetry in $3+1$ dimensions.

In this letter we show that the central charge of Beisert and the central charge of  the Coulomb branch of the ${\cal N}=4 $ SYM are really the same object. More precisely, the central charge of the Coulomb branch of ${\cal N}=4 $ SYM appears as a limit of the central charge of Beisert. Although this identification of the two central charges has been suggested before \cite{Berenstein:2007zf}, the arguments there are heuristic. In this paper we can make the argument very precise.
In order to do this, one has to show that there is an open string version of the central charge of Beisert, realized with appropriate boundary conditions,  which is exactly realized as the Coulomb branch of the ${\cal N}=4 $ SYM theory dual theory. The main part of the discussion is concerned with taking the flat space limit appropriately, so that even though all states in the ${\cal N}=4 $ SYM have a confined central charge on the sphere, one can argue that the compensating charges for a particular state  become hidden at infinity and are essentially decoupled from each other. This way,  one can ignore the interactions between them and treat them in isolation, where the central charge becomes indispensable.

\section{The one loop open spin chain and the central charge}

We want to make the central charge of Beisert physical in the gravity dual theory, rather than confined.   We will thus require the central charge to be sourced by D-branes, seeing as the spin chain model is the dual description of fundamental strings on the bulk geometry. We want the central charge to be carried by a D-brane object that preserves the same amount of supersymmetry as the ferromagnetic ground state of the spin chain. Under those conditions the unbroken symmetry coincides with one studied by Beisert. The ferromagnetic ground state preserves half the supersymmetries. The corresponding half BPS configurations in field theory are built from a single complex scalar degree of freedom which we call $Z$ (see \cite{Berenstein:2002jq,Berenstein:2013md} for conventions).

The D-branes that preserve this subgroup of the symmetry are provided by giant gravitons \cite{McGreevy:2000cw} and the so called dual giant gravitons \cite{Grisaru:2000zn, Hashimoto:2000zp}. Both of these correspond to D3-brane configurations in the dual $AdS_5\times S^5$ geometry, and their field theory duals are known \cite{Balasubramanian:2001nh, Corley:2001zk}. They are generally written in terms of Young tableaux made of the field $Z$.    The formalism to attach open strings to the D-branes in the dual field theory was argued for in \cite{Balasubramanian:2004nb}, which again uses decorated young tableaux to describe the states. 
The formalism makes apparent that the D-branes are spatially compact: one can not put a single string between two of them. After all, the endpoints of the strings are charged with respect to the gauge degrees of freedom on each such D-brane. Extra `return' strings are always required to cancel the charge. Eventually we want to argue that we can hide these strings at infinity, for any value of $g_{YM}^2N$ by taking an appropriate scaling limit.

The one loop computations for the open spin chain Hamiltonian in the $SU(2)$ sector depend on if we use regular giant gravitons, or dual giant gravitons \cite{de Mello Koch:2007uv} (see also \cite{Bekker:2007ea}). 
They can be written as follows
\begin{equation}
H_{giant} = g_{YM}^2 N\left[ (B-a_1^\dagger) (B^\dagger- a_1)+ (a^\dagger_1-a^\dagger_{2})( a_1-a_{2})+\dots\right]\label{eq:hgiant}
\end{equation}
where $B,B^\dagger $ are realized as a truncated harmonic oscillator algebra. These are also available at two loop order \cite{Berenstein:2013eya}.
 This means $B^\dagger \ket n = N^{-1/2} \sqrt{n+1} \ket{n+1}$, but the range of $n$ is restricted from above by $N$, the rank of the gauge group, plus corrections of order one depending on how many giant gravitons there are. Or similarly,
\begin{equation}
H_{dual giant} = g_{YM}^2 N\left[ (A^\dagger-a_1^\dagger) (A- a_1)+(a^\dagger_1-a^\dagger_{2})( a_1-a_{2})+\dots\right] \label{eq:hdgiant}
\end{equation}
where again $A^\dagger \ket n =  N^{-1/2} \sqrt{n+1} \ket {n+1}$ where the range of $n$ is bounded below by $N$ (also with corrections of order one depending on how many giant gravitons there are). This repackages various square roots that appear in the computation in terms of the algebra of raising and lowering operators for a harmonic oscillator.
Our notation is such that $n=N-b_0$, or $n=N+a_0$ in the notation of \cite{de Mello Koch:2007uv}, where this was explicitly computed, but not written in terms of ordinary harmonic oscillator algebras. The oscillators $a_i$ commute with each other, but they realize a Cuntz algebra \cite{Berenstein:2005fa}. The factors of $N^{-1/2}$ in $A, B$ are convenient in order to get all the factors of $N$ outside the 1-loop  effective Hamiltonian. Sets with many open strings and giants can be analyzed in this formalism (see \cite{Koch:2011hb} and references therein).

In the works  \cite{Berenstein:2013md,Berenstein:2013eya,Berenstein:2014isa}, the operators $B, B^\dagger$ were effectively replaced by c-numbers that describe the collective coordinate of the giant gravitons. These collective coordinates live on a disk in the complex plane of radius one. The factors of $N^{-1/2}$ on the oscillators indicate that the corresponding collective coordinate populates preferentially the values of $n\simeq O(N)$. A simple way to understand what this means is to assume that approximately
\begin{equation}
B \ket \xi \simeq \xi\ket \xi
\end{equation}
so that $\ket{\xi}$ is like a coherent state of a harmonic oscillator. Solving these equations (without assuming the truncations) we find that
\begin{equation}
\ket \xi \simeq \sum \frac{(\sqrt N\xi) ^n}{\sqrt{n!}} \ket n
\end{equation}
and then we impose the truncation $\ket n=0$ for $n>N$. This is a rewriting  of the results in \cite{Berenstein:2013md} in this harmonic oscillator language. In that work the starting point was completely different and the harmonic oscillator language was derived after the coherent states were postulated.  The truncation means that we can only trust the coherent states so long as $|\xi|<1$, and one can get near  the edge to about order $N^{-1/2}$ depending on the error one is willing to tolerate. This is exponentially suppressed on the distance to the edge of the disk. This point of view realizes very explicitly the holes in the free fermion picture droplet of the Half-BPS states \cite{Berenstein:2004kk}.

We now want to complete the picture above for the coherent states of the fermions in the fermion droplet picture, rather than the holes. This is, for  the dual giant states. The answer is pretty straightforward. We want to convert the operators $A, A^\dagger$ to c-numbers. We do this by solving the following equation
\begin{equation}
A\ket {\cal Z}={\cal Z} \ket {\cal Z}
\end{equation}
The solution is 
\begin{equation}
\ket {{\cal Z}}  \simeq \sum \frac{(\sqrt N {{\cal Z}} ) ^n}{\sqrt{n!}} \ket n
\end{equation}
and the only difference with a regular harmonic oscillator coherent state is that we need to truncate $n\geq N$.
Again, just as in \cite{Berenstein:2013md}, one can compute the norm of the state
\begin{equation}
\braket{\CZ}\CZ= \sum_{n\geq N} \frac{|N\CZ|^n}{n!}\simeq \exp( |N\CZ|^2)
\end{equation}
and the approximation to the norm of a regular coherent state of the harmonic oscillator (the right hand side) is good only if $|\CZ|>1+O(1/\sqrt  N)$. This is, the coherent state we find is very similar to a coherent state of an ordinary harmonic oscillator only so long as the collective coordinate $\CZ$ lives on the complex plane with the unit circle removed. It is easy to see that the  energy (above the ground state) is then 
\begin{equation}
\Delta= \vev{N A^\dagger A-N}=\vev{n-N}=N(|\CZ|^2-1).
\end{equation}
The factor of $N$ in front of $|\CZ|^2-1$ can be interpreted as a tension, and usual large $N$ counting indicates  that this corresponds to a D-brane state \cite{Berenstein:2013md}.
Here we use the convention that the energy on the sphere is related to the dilatation operator of the Euclidean field theory, which is why we called it $\Delta$. 
Notice that this structure of coherent states was conjectured also in \cite{Caldarelli:2004ig}.

Once we replace the $A, A^\dagger$ operators by c-numbers, the giant gravitons and dual giant gravitons appear essentially on the same 
footing on the boundary conditions. In the works \cite{Berenstein:2013eya,Berenstein:2014isa} it was argued that the Beisert central charge \cite{Beisert:2005tm} for an open string  state between giants was exactly 
$\xi-\tilde \xi$. Now we can get $ \CZ -\tilde \CZ$ for open strings between two dual giant gravitons or all possible combinations of 
dual-giant and giant states on both ends. What determines if we have a giant or dual giant is if the collective coordinate appearing in the spin chain is inside the unit radius disk, or outside of it. Given such a state, on a spin chain with $k$ sites, meaning there are $k+1$ units of $SO(4)$ charge, we expect an energy of the form
\begin{equation}
\Delta-J = \sqrt{(k+1)^2+ \frac{g_{YM}^2 N }{4\pi^2}|\xi -\tilde \xi|^2}
\end{equation} 
This is a generalization of the giant magnon bound state dispersion relation \cite{Dorey:2006dq} (see \cite{Berenstein:2014isa} for the choice of conventions). We can clearly now replace $\xi$ or $\tilde \xi$ by $\CZ$ or $\tilde \CZ$ or both. This is essentially identical to a relativistic dispersion relation
\begin{equation}
E\simeq \sqrt{p^2 +m^2}
\end{equation}
where $m\propto |\xi-\tilde \xi|$ and $p\propto k$. In this setup $k$ is quantized and it corresponds to the charge under the $SO(4)$ R-charge of the theory, but if one can take $k\to \infty$ appropriately and we rescale it, we can replace it by a continuous variable. Surprisingly, this is also argued to be the correct dispersion relation by doing a calculation that assumes we are in the Coulomb branch of ${\cal N}=4 $ SYM \cite{Berenstein:2007zf} in a saddle point limit (following ideas in \cite{Berenstein:2005jq}), but where now $k$ plays the role of an angular momentum on the boundary $S^3$. If we think of these as open strings ending on  D3-brane, the BPS states corresponding to W-bosons can only carry momentum along the directions of the brane, but not also transverse to it. Both types of states can not be BPS simultaneously. This presents a puzzle that we will resolve later in the paper.

\section{Dual giant gravitons and the Coulomb branch}

As noticed in \cite{Hashimoto:2000zp,Berenstein:2004kk}, a single dual giant graviton state corresponds to a classical symmetry breaking pattern $SU(N)\to SU(N-1)\times U(1)$.
When we quantize the classical physics and include the quantum effects,  the truncation of the lower levels for the oscillator pair $A, A^\dagger$ where the levels for $n<N$ are missing can be interpreted as a Fermi exclusion principle due to the free fermion description of the BPS states \cite{Berenstein:2004kk}.
As such, the oscillator pair $A^\dagger, A$ can be thought of as being the oscillator degrees of freedom of the eigenvalues of the field $Z$ directly and don't need any correction. With these conventions $Z$ acts as a raising operator $Z\simeq \diag ( \sqrt N A_i^\dagger+\dots)$.
 To do this we have a diagonal ansatz for $Z$, where the eigenvalues are fermionic degrees of freedom in a harmonic oscillator. Hence they are represented by an ordinary algebra of states in a harmonic oscillator. To describe physical states, one needs to impose the Fermi exclusion principle by hand: writing completely antisymmetric wave functions on the Hilbert space of states. 
  The lowering operator is in the conjugate field $\bar Z\simeq \diag \sqrt N A_i+\dots$. From here, the expectation values of $\vev{\Tr(\bar Z^k)}= \sum_i(\sqrt{N} \CZ_i(t))^k$ can be read directly from the collective coordinates if we have many such giants. We thus find that the c-numbers $\CZ(t)$ can be interpreted as the classical field solutions for  the equations of motion of the field $Z$ itself. 
  The only caveat in interpreting this immediately as the Coulomb brach is that the $Z$ depend on time as $\CZ(t)\simeq \CZ(0)\exp(it)$, and we find ourselves in a compact space: this is the correct result for field theory on an $S^3\times \BR$ geometry, not infinite flat space. 

There is a simple way to recover the Coulomb branch: take a limit where the mass scales of any interesting process is much higher than the  scale determining the radius of the sphere $S^3$. At the same time, we want only degrees of freedom associated to field theory in the Coulomb branch of an $U(2)$ gauge theory to survive,  but  strings should. This is, we want to take a limit where $|\sqrt N \CZ R| \to \infty$. Here, $R$ is the radius of the sphere (it has been set to one in the convention so far). In this limit, the vacuum expectation value of the field $Z$ is going to infinity. We want to change units so that the value of the field is of order one. This means we can think equivalently as the field being fixed in size, but taking $R\to \infty$ instead.
In such a situation, we are also rescaling the time scale for a process to be that associated to the scale of the expectation value of the field, rather than the  time that it takes to go around the sphere on the $S^3\times \BR$ geometry. Under such conditions, the times scales are short, and the effective frequency that controls the evolution of the field $Z$ goes to zero. We can argue that we are therefore in a static situation for the field, on the Coulomb branch. For such short timescales, an excitation at one location of the sphere has no time to communicate to a return string that is located close to the opposite side of the sphere. The return string is essentially hidden at infinity after the rescaling. The local excitations can be addressed in isolation ignoring the global charge constraint. Now, the central charge has become completely physical.

If we have two such eigenvalues with a collective coordinate turned on, we have broken $SU(N)\to SU(N-2)\times U(1)\times U(1)$.
A string stretching between the two dual giants will have a Beisert central charge $\CZ_1-\CZ_2$, and it can be thought of as an off-diagonal  field of SYM charged under both $U(1)$. The mass we associate with it is 
$g_{YM} \sqrt N (\CZ_1-\CZ_2)$, which we want to be large (much larger than $R^{-1}$, even for arbitrarily small $g_{YM} N$). However, the $SU(N-2)$ still present suggests that we have to think about the problem in terms of strong coupling physics. To decouple the degrees of freedom charged under both $U(1)\times U(1)$ from the rest , we want the other off-diagonal degrees of freedom to be much more massive. This is, we want to take the limit in such a way that 
\begin{equation}
\frac{|(\CZ_1-\CZ_2)|}{|\CZ_1|} \to 0 \label{eq:limit}
\end{equation}
Moreover, we still need to get rid of the possibility of excited strings between the dual giants.  This requires taking this limit in a more careful fashion. We want the two dual giants to be ``close" to each other in the AdS geometry. Closer than the local string scale at the location of the branes. This requires translating the $\CZ$ coordinates to global $AdS$ coordinates. Following \cite{Hashimoto:2000zp}, if we have an AdS metric of the form
\begin{eqnarray}
ds^2 &=& -\cosh^2 \varrho dt^2 +d\varrho^2+\sinh^2 \varrho d\Omega_3^2\\
&=& \frac 1 {\cos^2 \rho}(-dt^2+ d\rho^2 +\sin^2\rho d\Omega_3^2)
\end{eqnarray}
a dual giant graviton with momentum $L$ is located at a radial position
\begin{equation}
\tan \rho= \sinh \varrho = \sqrt{L/N}= \sqrt{|\CZ|^2 -1}
\end{equation}
Here we are using the fact that $\Delta = L = N |\CZ|^2 -N$.
This means that the redshift factor is $\cosh \varrho= |\CZ|\simeq |\CZ_1|\simeq| \CZ_2|$, and for the branes to be closer than the string scale, the energy of a string suspended between them needs to be lower than the local string length. This means that we want to take the limit such that
\begin{equation}
1 << g_{YM} \sqrt N |\CZ_1-\CZ_2| << \ell^{-1}_s\cosh \varrho= (g_{YM}^2 N)^{1/4} |\CZ|
\end{equation} 
this only modifies equation \eqref{eq:limit} at strong t'Hooft coupling. Obviously both limits can be realized by first choosing $|\CZ_1-\CZ_2|$ sufficiently large, and then choosing $|\CZ|$ to be that much larger. Notice that in this limit the branes can also be considered to be static. This is because the proper velocity of the branes on a rest frame located at the $AdS$ position of the brane is $v_{proper} = 1/ \cosh(\varrho) \to 0$, an effect due to the blueshift of proper time.

In this limit, at energies of order $g_{YM} \sqrt N |\CZ_1-\CZ_2| $ there is no way to excite stringy states that are not low energy W-bosons. Similarly, we can argue that there is no back reaction so long as the energy of these $W$ bosons is much smaller than the energy of the brane $g_{YM} \sqrt N |\CZ_1-\CZ_2|<<L\simeq N|\CZ|^2$. Since the right hand side grows quadratically in $|\CZ|$, there is always a sufficiently large value of $\CZ$ for which the energy of the W-bosons is subleading, even at finite values of $g_{YM}$. 
Any calculation done in this double scaling limit for dual giants can be considered to be exactly in the Coulomb branch of ${\cal N}=4$ SYM in flat space. Here we now see the identification of the central charge of Beisert as being identical in value to the central charge of ${\cal N}=4 $SYM, even at finite coupling in the appropriate limit. 

\section{Walls of  stability}

The next question that needs to be asked is if the BPS states with given quantum numbers exist or if they don't. We will address this explicitly in the $SU(2)$ sector at weak coupling, and will conjecture what the rules for the other states are. The key observation is that the ground state of the spin chain Hamiltonian described by \eqref{eq:hgiant} and \eqref{eq:hdgiant} can be computed in the presence of coherent states for giants or dual giants, along the lines of \cite{Berenstein:2013eya}. 
\begin{equation}
H_{1-loop} = g_{YM}^2 N\left[ (w^*-a_1^\dagger) (w- a_1)+ \sum_{i=1}^{k-1} (a^\dagger_i-a^\dagger_{i+1})( a_i-a_{i+1})+
(a_k^\dagger -\tilde w^*) (a_k-\tilde w)\right]\label{eq:hamc}
\end{equation}
The idea is to introduce coherent states for the Cuntz oscillators $a_i$, such that
\begin{equation}
a_i \ket {z_i} = z_i \ket{z_i} 
\end{equation}

Substituting in the Hamiltonian, the energy becomes a simple quadratic form
\begin{equation}
\vev{H_{1-loop}} = g_{YM}^2 N \left[(w^*-z_1^*)(w-z_1) +\sum_{i=1}^{k-1} |z_i-z_{i+1}|^2+(z^*_k -\tilde w^*) (z_k-\tilde w)\right]
\end{equation}
which can also be written as
\begin{equation}
\vev{H_{1-loop}} = g_{YM}^2 N \sum_{i=0}^{k-1} |z_i-z_{i+1}|^2
\end{equation}
with $z_0= w$ and $z_{k+1}= \tilde w$. The energy is minimized if $z_i-z_{i+1}= z_{i-1}-z_i$ for all $i=1, \dots k$. This actually corresponds to an exact eigenstate of the Hamiltonian, with the minimal value of the energy. The values of $z_\ell$ are then determined readily as
\begin{equation}
z_\ell = w + \frac{\tilde w -w}{k+1} \ell
\end{equation}
and they interpolate linearly between $w, \tilde w$. The central charge of the state would be characterized by $w-\tilde w$.

In terms of an orthonormal basis of occupation number, the coherent states for the Cuntz oscillator are given by
\begin{equation}
\ket z_i \propto \sum_{s=0}^\infty z_i^ s \ket s
\end{equation}
The state has finite norm only if $\sum |z_i|^{2s} <\infty$. 
A would-be ground state is therefore normalizable only if the $|z_i|<1$ for all $i=1, \dots, k$. If we fix $w, \tilde w$, it is clear that when both $w, \tilde w$ are inside the unit disk (including the edge), that all $z_i$ are inside the unit disk. However, if one or both of $w,\tilde w$ are outside the unit disk, there always exists a sufficiently large $k$ for which at least one $|z_\ell|>1$ (usually many). How many such states have that property for a fixed $k$ depend on the positions of $w,\tilde w$ in the complex plane. As we move them (by choosing different giant gravitons or dual giant gravitons), a $z_i$ might cross from being inside the unit disk to going outside it and the state disappears. It is interesting to characterize that motion in terms of a physical observable, rather than the norm of the coherent state which can always be chosen to be equal to one before taking the limit.

The simplest way to do so is to compute the expectation value of the occupation number operator for a site $i$ on such a coherent state,
\begin{equation}
\vev{\hat N_i }= N_z \sum_{s=0}^\infty s  |z_i|^{2s} = \frac{|z_i|^2}{(1-|z_i|^2)}
\end{equation}
This evaluates to a number greater than zero if $|z_i|<1$, and it becomes infinity when $|z_i|\to 1$. Beyond that, the value can be analytically continued for $|z_i|>1$, but it is negative in value, which can not be possible (the number operator is  positive on the Hilbert space of states). The way this happens is that a quantum number for a state becomes infinitely large. This is reminiscent of other setups where BPS states are lost when they become physically infinitely large \cite{Denef:2000nb}, where a naive analysis would allow an analytic continuation that makes the state unphysical by having a negative size. The condition of positive size then becomes a condition that determines if the corresponding state is stable or unstable. What should happen when the system stops having a BPS ground state is that supersymmetry is broken. It is reasonable to expect that in this case the spectrum of the spin chain becomes continuous , along the lines of \cite{Berenstein:2006qk}. The transition that should get rid of the BPS state is that the gap from the ground state to the continuum closes. In the full ${\cal N}=4 $ SYM the spectrum of energies will be eventually discrete. This only comes about from considering back reaction of the D-branes in which the strings end and this is subleading when counting powers of $N$.  The stability question of these magnon is also very reminiscent of the stability problem for 'twisted magnons' in ${\cal N}=2 $ theories \cite{Gadde:2010ku}.

A rather important question is what to do with the $SL(2)$ sector, which is the sector that gives rise  to the results in \cite{Berenstein:2007zf}.
The simplest way to think about it is to go to the plane wave limit \cite{Berenstein:2002jq}. In that limit the $SO(4)$ of R-charge and the $SO(4)$ of rotations on the $S^3$ appear on the same footing. Indeed, even the free fermion picture -- as realized in supergravity-- simplifies \cite{Lin:2004nb}, and the supergravity theory in this limit treats holes and fermions exactly on the same footing. The natural conjecture is that what we have been able to show in the $SU(2)$ sector should be copied to the $SL(2)$ sector.
 As we have seen, the stability of BPS states are determined by the physics near the edge of the unit circle, and when we zoom on this edge we get the plane wave limit. For two giants that are holes, we get this sequence of intermediate values of $z_i$ on the filled area of the droplet. It should be the case that for two giants that are fermions, (dual giants) we get a similar sequence of intermediate values of $z_i$ describing the state on the unfilled area of the droplet, even though it is not clear how to interpret them in terms of coherent states of the Cuntz oscillator. Such a system should describe the BPS (bound) states in the $SL(2)$ sector. The classical string solutions associated to such BPS states should be very similar to the giant magnons \cite{Hofman:2006xt} and their dyonic counterparts \cite{Chen:2006gea}, but they should reside in  $AdS$ space times a circle rather than the sphere times time.  It is likely that such states can be related to the spike solutions in the string sigma model \cite{Kruczenski:2004wg}, but they do not have a logarithmic correction to the energy in terms of the spin like those that show up in purely AdS solutions
\cite{Gubser:2002tv}. This seems to be related to the intuition that spinning them faster does not seem to get them closer to the boundary.
The study of such string solutions is beyond the scope of the present paper.
 
 These states will only be bound because of the presence of the central charge and the finite length of the string.  After all, the naive interaction in the sigma model  on departing from the plane wave limit between $SL(2)$ impurities  is repulsive, due to the $AdS$ being negatively curved, whereas in contrast the five sphere is positively curved and the $SU(2)$ sector interaction is attractive.
 When these other  BPS states in the $SL(2)$ sector  cross into the filled area they disappear. What this means is that the states that have the correct energy in  \cite{Berenstein:2007zf} do not really exist as normalizable bound states. They exist only in the Coulomb branch without taking into account the non-trivial quantum corrections from the fermion droplet.

\acknowledgements 

The author would like to thank N. Dorey for discussions and E. Dzienkowski for many discussions and comments on the manuscript.
Work  supported in part by the department of Energy under grant {DE-SC} 0011702. The research leading to these results has received funding from the European Research Council under the European Community's Seventh Framework Programme (FP7/2007-2013) / ERC grant agreement no. [247252].

\end{document}